\documentclass[9pt]{extarticle}
\usepackage[utf8]{inputenc}

\usepackage{amsmath}
\usepackage{amssymb}
\usepackage{authblk}
\usepackage[textwidth=14.0cm]{geometry}
\usepackage{graphicx}
\usepackage{physics}
\usepackage{caption}
\usepackage{latexsym}
\usepackage[style=phys, biblabel=brackets]{biblatex}
\bibliography{ref.bib}
\usepackage{textgreek}
\newcommand{\deri}[2]{\frac{\partial #1}{\partial #2}}

\renewcommand{\vec}[1]{\boldsymbol{#1}}
\newcommand{\mat}[1]{\boldsymbol{#1}}
\newcommand{\erfc}{\text{erfc}}
\renewcommand{\erf}{\text{erf}}
\renewcommand{\order}[1]{\mathcal{O}\left(#1\right)}
\DeclareMathSymbol{\shortminus}{\mathbin}{AMSa}{"39}

\newcommand{\heaviside}[1]{\theta\!\left(#1\right)}
\newcommand{\changed}[1]{#1}
\newcommand{\permativity}{}
\newcommand{\tave}[1]{\left\langle\!\left\langle#1\right\rangle\!\right\rangle}

\renewenvironment{abstract}
 {\small
  \begin{center}
  \bfseries \abstractname\vspace{-.5em}\vspace{0pt}
  \end{center}
  \list{}{%
    \setlength{\leftmargin}{0pt}% <---------- CHANGE HERE
    \setlength{\rightmargin}{0pt}%
  }%
  \item\relax}
 {\endlist}

\title{Development of a new quantum trajectory molecular dynamics framework}
\author[1]{Pontus~Svensson}
\author[1]{Thomas~Campbell}
\author[2]{Frank~Graziani}
\author[3]{Zhandos~Moldabekov}
\author[4]{Ningyi~Lyu}
\author[4,5]{Victor~S.~Batista}
\author[6]{Scott~Richardson}
\author[1,7]{Sam~M.~Vinko}
\author[1]{Gianluca~Gregori}

\affil[1]{Department of Physics, University of Oxford, Parks Road, Oxford OX1 3PU, UK}
\affil[2]{Lawrence Livermore National Laboratory, Livermore, CA 94550, USA}
\affil[3]{Center of Advanced Systems Understanding (CASUS), Helmholtz-Zentrum Dresden-Rossendorf, Untermarkt 20, 02826 Görlitz, Germany}
\affil[4]{Department of Chemistry, Yale University, New Haven, CT 06520, USA}
\affil[5]{Yale Quantum Institute, Yale University, New Haven, CT 06511, USA}
\affil[6]{AWE, Aldermaston, Reading, Berkshire RG7 4PR, UK}
\affil[7]{Central Laser Facility, STFC Rutherford Appleton Laboratory, Didcot OX11 0QX, UK}

\date{\today}

\begin{document}

\maketitle

\begin{abstract}
\noindent An extension to the wave packet description of quantum plasmas is presented, where the wave packet can be elongated in arbitrary directions. A generalised Ewald summation is constructed for the wave packet models accounting for long-range Coulomb interactions and fermionic effects are approximated by purpose-built Pauli potentials, self-consistent with the wave packets used. We demonstrate its numerical implementation with good parallel support and close to linear scaling in particle number, used for comparisons with the more common wave packet employing isotropic states. Ground state and thermal properties are compared between the models with differences occurring primarily in the electronic subsystem. Especially, the electrical conductivity of dense hydrogen is investigated where a $15\%$ increase in DC conductivity can be seen in our wave packet model compared to other models.
\end{abstract}

%\input{text/1_Introduction}
%\input{text/2_Theory}
%\input{text/3_Numerics}
%\input{text/4_Test}
%\input{text/6_Conclusion}

%\appendix
%\input{text/A_Decomposition}
%\input{text/B_Ewald}
%\input{text/C_Ewald_Pauli}
%\input{text/D_Temperature}
%\input{text/E_Friction}

\section{Introduction}%
The establishment of high power lasers facilities during the last decades has been instrumental in the achievements towards inertial confinement fusion (ICF)\,\cite{nuckolls1972laser,betti2016inertial,abu2022lawson}, but also for the creation of high-density and high-temperature conditions\,\cite{smith2018equation} otherwise only found in astrophysical objects\,\cite{falk2018experimental,frank2019exoplanets}. Furthermore, X-ray lasers are now able to reach complementary high-pressure regions in phase space\,\cite{levy2015creation,fletcher2015ultrabright}. One of the exotic states now accessible is warm dense matter (WDM) which exists in gas giants\,\cite{schlanges1995plasma,bezkrovniy2004monte,vorberger2007hydrogen,lorenzen2014progress,militzer2021first}, brown\,\cite{saumon1992role} and white dwarf stars\,\cite{chabrier1993quantum,chabrier2000cooling}, the crust of neutron stars\,\cite{haensel2007neutron,daligault2009electron}, and during the compression of an ICF capsule\,\cite{hu2018review}. Warm dense matter is a strongly coupled quantum plasma, with ions moving in a partially degenerate electron fluid with kinetic energy comparable to the ion-ion interaction energy\,\cite{bonitz2020ab}. Consequently, WDM inherits properties from both condensed matter systems and classical plasmas, a challenging combination to model. Various computational techniques are commonly used to describe the system, yielding similar thermodynamic\,\cite{gaffney2018review} and acoustic properties\,\cite{davis2020ion,yao2021reduced}, although dynamic properties differ by orders of magnitude\,\cite{grabowski2020review}. These uncertainties limit our understanding of, for example, the Jovian interior\,\cite{wahl2017comparing}, or the modelling of ICF implosions\,\cite{hu2015impact}.  

The three main complications in modeling WDM are electron degeneracy, strong ion correlations and the separation in timescales between the electron and ion dynamics. A full solution would require a quantum mechanical treatment of the electrons, resolving electron dynamics while considering phenomena on the ion time scale. Consequently, explicit models of ionic motion span a wide range of theories, including classical systems with effective ion-ion interactions\,\cite{ichimaru1982strongly,mithen2011extent,kahlert2020thermodynamic}, classical electrons with effective quantum statistical potentials (QSP)\,\cite{minoo1981temperature,hansen1983thermal,glosli2008molecular}, Bohmian mechanics\,\cite{larder2019fast}, density functional theory molecular dynamics (DFT-MD) using both orbital-free\,\cite{lambert2007properties,white2013orbital} and Kohn-Sham\,\cite{collins1995quantum,ruter2014ab,lorenzen2014progress} DFT-variants, phenomenological quantum hydrodynamics based on DFT-functionals\,\cite{moldabekov2018theoretical,michta2020quantum,moldabekov2022towards}, time-dependent DFT\,\cite{baczewski2016x,magyar2016stopping,frydrych2020demonstration}, and quantum Monte-Carlo and path integral Monte-Carlo\,\cite{militzer2000path,hu2011first,militzer2021first} approaches. Coarse-grained models with effective interactions are fundamentally based on reconstructing some equilibrium property, the choice of which is arbitrary and limited to a specific thermodynamic condition, whereas experimental realisations are commonly non-stationary\,\cite{chapman2011analysis,zastrau2014resolving,white2014electron,plagemann2015ab,clerouin2015evidence,levy2015creation}. Furthermore, models rooted in the Born-Oppenheimer approximation -- where the electrons are treated adiabatically -- e.g. DFT-MD cannot capture a dynamic electron response, believed to be important for the description of dynamic properties such as some transport coefficients\,\cite{mabey2017strong}, stopping power\,\cite{michta2020quantum} and energy transfer between the electronic and ionic subsystems. However, time-dependent approaches are computationally costly, and are typically limited in terms of particle numbers and time scales of studied phenomena. 

Wave packet molecular dynamics (WPMD)~\cite{feldmeier2000molecular,grabowski2014review} is a family of models in which the electron dynamics are computed explicitly, whiles simulating hundreds to thousands of particles over ionic time scales. This is made possible by restricting the wave function of each electron to a parameterised functional form. We present an extension to existing wave packet formulations -- applicable to the WDM regime -- in which the wave packets can be elongated in arbitrary directions. The model accounts for the long-range behaviour of electrostatic interactions and of fermionic properties by effective Pauli interactions, while implemented within the scalable molecular dynamics framework \texttt{LAMMPS}\,\cite{plimpton1995fast} to treat systems with thousands of particles. 

In the following section the theoretical model is described, after which section \ref{sec:numerics} outlines the numerical details and performance of the implementation. The model is compared with other computational techniques in section \ref{sec:validation}, where we apply it to ground state and dynamic properties of a dense hydrogen plasma. We compute some structural and transport properties, which are compared with an isotropic wave packet model. We conclude with a summary of our results. 
%
%\input{text/2_Theory}
% ------------------- Theory ---------------------------
\section{Theoretical Description}\label{sec:theory}
Originally proposed in the 1970s as an approximate solution to Schrödinger's equation~\cite{heller1975time,corbin1982semiclassical}, wave packet models can systematically be derived from variations of the action,
\begin{equation}
    \mathcal{S} = \int\!dt\, \bra{Q}  i\hbar \frac{d}{dt} - \hat{H} \ket{Q},
\end{equation}
where $\hat{H}$ is the system Hamiltonian and the state, ${\ket{Q} = \ket{Q(Q_{\mu})}}$, is restricted to some manifold, $\mathcal{M}$, defined by the adopted wave packets and parameterised by its parameters $Q_{\mu}$. The resulting time evolution reproduces the true quantum dynamics to the best of its ability being restricted to the manifold, $\mathcal{M}$, during short time scales of length $\delta t$. Concretely, it can be shown $\bra{\Delta(t, \delta t)}\ket{\Delta(t, \delta t)}$ is minimised to $\mathcal{O}(\delta t^3)$ where $\ket{\Delta(t, \delta t)} = \ket{\Psi(t + \delta t)} - \ket{Q(t + \delta t)}$ and $\ket{\Psi(t + \delta t)}$ is the true solution to Schrödinger's equation starting from $\ket{\Psi(t)} = \ket{Q(t)}$~\cite{feldmeier2000molecular}. The long-time evolution is constrained by appropriate conservation laws, most notably energy conservation\,\cite{broeckhove1989time}. 

In general, the equations of motion are quasi-Hamiltonian
\begin{equation}
    i\hbar \sum_{\nu} \mathcal{C}_{\mu\nu} \frac{d Q_{\nu}}{dt} = \deri{\mathcal{H}}{Q^*_{\mu}}
    \label{eq:full_wave_packet_dynamics}
\end{equation}
where 
\begin{equation}
    \mathcal{H} = \frac{\bra{Q} \hat{H} \ket{Q}}{\bra{Q}\ket{Q}} \equiv \langle \hat{H} \rangle \quad \text{and} \quad \mathcal{C}_{\mu\nu} = \frac{\partial^2}{\partial Q_{\mu}^* \partial Q_{\nu}} \ln\Big(\bra{Q(Q_{\mu}^*)}\ket{Q(Q_{\nu})}\Big).
\end{equation}
Fermions are described by states anti-symmetric under exchange and Slater-determinants have been considered in Refs.\,\cite{knaup2003wave,jakob2007wave,jakob2009wave}. However, this approach scales unfavourably with particle number $N$. Instead, here we employ a product state
\begin{equation}
    \ket{Q} = \ket{q_{1}} \otimes \ket{q_{2}} \otimes \dots \otimes \ket{q_{N}}
\end{equation}
of single-particle orbitals, $\ket{q_i}$, and $\otimes$ is the tensor product. Exchange effects are approximated by Pauli potentials of the type first introduced by Klakow et~al.\,\cite{klakow1994hydrogen,klakow1994semiclassical}. This structure simplifies $\mathcal{C}_{\mu\nu}$, which becomes block diagonal, and the orbitals $\ket{q_i}$ only couple through the energy $\mathcal{H}$.
    
\subsection{Wave Packets}
    The choice of wave packet shape is central to the model, dictating the states that can be described\changed{\,\cite{murillo2003self}}. Most commonly, isotropic gaussians are utilised -- primarily motivated by computational ease -- yet other variants exist, see Ref.\,\cite{grabowski2014review} and references therein. To account for local gradients, an anisotropic wave packets form is introduced 
    \begin{equation}
        \bra{\vec{x}}\ket{q_i} = \left(\left(2\pi\right)^3 \det\left(\mat{\Sigma}_i\right) \right)^{- 1/4} \times \exp\left[ - \vec{\xi}_i \left(  \frac{1}{4} \mat{\Sigma}_i^{-1} - \frac{i}{\hbar}\mat{\Pi}_i \right) \vec{\xi}_i + \frac{i}{\hbar} \vec{p}_i^\intercal \vec{\xi}_i \right],
        \label{eq:anisotropic_state}
    \end{equation}
    where $\vec{\xi}_i = \vec{x}  - \vec{r}_i$. The wave packet is parameterised by 18 degrees of freedom, the position $\vec{r}_i$, momentum $\vec{p}_i$ and two symmetrical \changed{$3 \times 3$} matrices, $\mat{\Sigma}_i$, describing the elongation and orientation, and $\mat{\Pi}_i$ the associated momentum to $\mat{\Sigma}_i$. A similar type of wave packet has been treated previously, describing molecular binding in water molecules\,\cite{ono2012semiquantal}, and is generally believed to improve the description of molecular states\,\cite{grabowski2014review}.   
    
    The equations of motion for the functional form \eqref{eq:anisotropic_state} have a classical-looking structure, for the ``classical'' degrees of freedom\,\cite{ono2012semiquantal}
    \begin{subequations}
        \begin{equation}
            \frac{d\vec{r}_i}{dt} = \deri{\mathcal{H}}{\vec{p}_i}, \hspace{0.5cm} \frac{d\vec{p}_i}{dt} = - \deri{\mathcal{H}}{\vec{r}_i}
        \end{equation}
        and the ``internal'' dynamics of the wave packet follow as 
        \begin{equation}%
            \changed{\frac{d}{dt}\Sigma_{i\alpha\beta} = \tau_{\alpha \beta} \deri{\mathcal{H}}{\Pi_{i \alpha \beta}}, \hspace{0.5cm} \frac{d}{dt}\Pi_{i \alpha \beta} = - \tau_{\alpha\beta} \deri{\mathcal{H}}{\Sigma_{i \alpha \beta}},}
            \label{eq:eom_internal}
        \end{equation}%
        \label{eq:EoM}%
    \end{subequations}
    \changed{where $\Sigma_{i\alpha\beta} = \left(\mat{\Sigma}_i \right)_{\alpha\beta}$ and $\Pi_{i\alpha\beta} = \left(\mat{\Pi}_i \right)_{\alpha\beta}$ are the components of the symmetric matrices. The pre-factor $\tau_{\alpha\beta}$ is unity if $\alpha = \beta$ and one half otherwise, which accounts for the symmetric structure of $\mat{\Sigma}_i$ and $\mat{\Pi}_i$ where $\Sigma_{i\alpha\beta}$ ($\Pi_{i\alpha\beta}$) and $\Sigma_{i\beta\alpha}$ ($\Pi_{i\beta\alpha}$) are treated as symbolically the same.} Specifically, we consider a charged system of classical ions, with position $\vec{R}_I$, momentum $\vec{P}_I$, charge $Z_Ie$ and mass $M_I$, and quantum electrons with position $\hat{\vec{x}}_i$ and momentum $\hat{\vec{p}}_i$ operators as well as charge $-e$ and mass $m$. The system is described by the Hamiltonian
    \begin{equation}
        \hat{H} = \sum_{I} \frac{\vec{P}_I^2}{2M_I} + \sum_{I < J} \frac{Z_I Z_J e^2}{\permativity |\vec{R}_I - \vec{R}_J|} + \sum_{i} \frac{\hat{\vec{p}}_i^2}{2m} + \sum_{i<j} \frac{e^2}{\permativity |\hat{\vec{x}}_i - \hat{\vec{x}}_j|} - \sum_{i} \sum_{I} \frac{Z_I e^2}{\permativity |\hat{\vec{x}}_i - \vec{R}_I| },
    \end{equation}
    the state average of which is required for the time evolution. The average kinetic energy 
    \begin{equation}
        \left\langle \frac{\hat{\vec{p}}_i^2}{2m}\right\rangle = \frac{\vec{p}_i^2}{2m} + \frac{2}{m} \Tr\Big\{ \mat{\Pi}_i \mat{\Sigma}_i \mat{\Pi}_i \Big\} + \frac{\hbar^2}{8m} \Tr\Big\{ \mat{\Sigma}_i^{-1} \Big\}  
        \label{eq:kin}
    \end{equation}
    includes both a classical contribution and a part internal to the wave packet. The last term in equation \eqref{eq:kin} is the so-called shape-kinetic energy\,\cite{wyatt2006quantum}, which keeps $\mat{\Sigma}_i$ positive definite and the wave-packet well defined during the time evolution. The interaction terms have not been evaluated explicit and the following section is dedicated to the treatment of these terms.
    
\subsection{Generalised Ewald summation}\label{sec:ewald}
    Within molecular dynamics, it is desirable to truncate pair-interactions at some distance such that the computation formally scales as  $\order{N}$\,\cite{rapaport2004art}. However, in our case the electrostatic interaction is long-range\,\cite{de1980simulation} and it is beneficial to perform the split\,\cite{deserno1998mesh,kolafa1992cutoff}
    \begin{equation}
        \frac{1}{r} = \frac{\erfc\left(gr\right)}{r} + \frac{\erf\left(gr\right)}{r},
        \label{eq:ewald_split}
    \end{equation}
    chosen so that the first term can be truncated at a distance of order $g^{-1}$, while the second term is regular as $r \rightarrow 0$ and efficiently evaluated in Fourier space. The Ewald parameter $g$ is chosen to optimise performance. Below we present a self-consistent treatment of both terms, as the long-range part has only been mentioned once for isotropic wave packets\,\cite{knaup1999wave} and is commonly neglected.
        
    \paragraph{Short-range forces:}
        In the case of a gaussian interaction kernel, the required state average can promptly be evaluated\,\cite{frantsuzov2004quantum,shigeta2013quantal}. Therefore, we construct a gaussian decomposition of the interaction kernel, 
        \begin{equation}
            \frac{\erfc\left(gr\right)}{r} \simeq \sum_{p} c_p e^{-\alpha_p r^2},
            \label{eq:decomposition}
        \end{equation}
        where the coefficients $c_p$ and $\alpha_p$ are fitting parameters. A robust numerical scheme to perform the decomposition is described in appendix \ref{app:decomposition}, where typically only 5-15 modes are required. By approximating the potential form, the notion of energy conservation is retained. We note that this is not the case for methods based on Taylor expansions\,\cite{ono2012semiquantal} or on direct numerical evaluations\,\cite{buch2002exploration}, due to either truncation errors or numerical noise.
        
    \paragraph{Long-range forces:}
        To limit surface effects in a finite size simulation, the simulation box is periodically repeated. Periodic images are included in accordance with the standard treatment of Ewald summation with particles positioned at $\vec{r}_i + L\vec{n}$ for all $\vec{n} \in \mathbb{Z}^3$ and $L$ being the length of a cubic simulation cell. The interaction energy is 
        \begin{equation}
            \sum_{i < j} \left\langle V(\vec{x}_{ij})\right\rangle \rightarrow \frac{1}{2}\sum_{i,j}' \sum_{\vec{n} \in \mathbb{Z}^3} \left\langle V(\vec{x}_{ij} \shortminus L\vec{n})\right\rangle  = \underbrace{\frac{1}{2} \sum_{i, j} \sum_{\vec{n} \in \mathbb{Z}^3} \left\langle V(\vec{x}_{ij} \shortminus L\vec{n})\right\rangle }_{\mathcal{E}_k} - \underbrace{\frac{1}{2}\sum_{i = j} \left\langle V(\vec{x}_{ij})\right\rangle }_{\mathcal{E}_s},
            \label{eq:ewald_summation}
        \end{equation}
        where $V$ is the long-range part of the Coulomb interaction in equation \eqref{eq:ewald_split}. The special case $i = j$ is excluded when $\vec{n} = 0$ (denoted by the primed sum) resulting in two distinct terms the main contribution $\mathcal{E}_k$ and the self-energy $\mathcal{E}_s$. In appendix \ref{app:ewald}, we evaluate $\mathcal{E}_k$ in reciprocal space to be 
        \begin{equation}
            \mathcal{E}_k = \frac{1}{2L^3} \sum_{\vec{k} \neq 0} \frac{4\pi}{k^2} e^{-k^2/4g^2} |\Tilde{\rho}_{\text{uc}}(\vec{k})|^2,
            \label{eq:long_main}
        \end{equation}
        where $\Tilde{\rho}_{\text{uc}}(\vec{k})$ is the charge density
        \begin{equation}
            \Tilde{\rho}_{\text{uc}}(\vec{k}) = e \sum_i Z_i e^{-i \vec{k} \cdot \vec{r}_i} e^{-\frac{1}{2} \vec{k}^\intercal \mat{\Sigma}_i \vec{k}},
            \label{eq:FT_density}
        \end{equation}
        and ${\vec{k} = 2\pi \vec{n} / L}$. Equation \eqref{eq:long_main} converges rapidly due to the exponential factor. The self-energy term, $\mathcal{E}_s$, is independent of $\vec{r}_i$ and does not influence the classical degrees of freedom directly. Although for distributed particles, it does depend on $\mat{\Sigma}_i$ and influences the internal dynamics. Based on the decomposition of the long-range interaction kernel (appendix \ref{app:decomposition}) the self-energy is evaluated in appendix \ref{app:ewald}. 
    
\subsection{Pauli interactions}\label{sec:pauli}
    The difference in the kinetic energy between a pair-wise anti-symmetrised state and the product state has often been used as a Pauli potential to correct for the fermionic structure of electrons\,\cite{klakow1994hydrogen,klakow1994semiclassical,knaup1999wave}. The electron force field model (eFF) introduced fitting parameters in the Pauli interaction to achieve stable bounds for elements with $Z \leq 6$\,\cite{su2007excited} and has been widely used with minor modifications\,\cite{su2009dynamics,jaramillo2011large,kim2011high,xiao2015non,ma2019extremely,davis2020ion,yao2021reduced,liu2021molecular}. Ref.\,\cite{angermeier2021investigation} considered exchange contributions to the interaction terms while including a correlation potential with a free parameter. This treatment of the Pauli interaction is extended here to anisotropic gaussian states.
    
    We construct the potential by considering two particles, $i$ and $j$, with the Hamiltonian
    \begin{subequations}
    \begin{equation}
        \hat{H}_2 = \frac{\hat{\vec{p}}^2_i + \hat{\vec{p}}^2_j}{2m} + \frac{e^2}{|\hat{\vec{x}}_i - \hat{\vec{x}}_j|} + V_{\text{bg}}\left( \hat{\vec{x}}_i, t \right) + V_{\text{bg}}\left( \hat{\vec{x}}_j, t \right)
    \end{equation}
    with a background interaction from all other particles in the system
    \begin{equation}
        V_{\text{bg}}\left( \vec{x}, t \right) = \!\!\sum_{k \neq i, j}e^2\int d^3\vec{x}_k\frac{|\bra{\vec{x}_k}\ket{q_k}|^2}{|\vec{x} - \vec{x}_k|} - \sum_{I} \frac{Z_I e^2}{|\vec{x} - \vec{R}_I|},
        \label{eq:pauli_bg}
    \end{equation}
    \end{subequations}
    where $k$ ($I$) runs over all other electrons (ions) in the system. The two-electron system can be characterised based on its spin structure either as a singlet or a triplet state, requiring either a symmetric or anti-symmetric spatial state  
    \begin{equation}
        \text{Singlet: } \ket{q_i}\otimes\ket{q_j} + \ket{q_j}\otimes\ket{q_i} \qquad \text{Triplet: } \ket{q_i}\otimes\ket{q_j} - \ket{q_j}\otimes \ket{q_i}
        \label{eq:single_triplet_space}
    \end{equation}
    written here in terms of single particle orbitals $\ket{q_i}$. 
    
    For equal spin particles only the spatially anti-symmetric state is allowed and the Pauli potential $\mathcal{V}_{ij}^{P}$ is the difference between $\langle \hat{H}_2 \rangle$ for the triplet and the product state $\ket{q_{i}} \otimes \ket{q_{j}}$. In the case of opposite spin particles, the spatial state depends on the spin structure, but along the lines of Ref.\,\cite{angermeier2021investigation} a correlation potential is introduced based on the singlet state, $\rho \mathcal{V}_{ij}^{C}$, multiplied by the parameter $\rho$. Therefore, the potentials are  
    \begin{equation}
        \mathcal{V}^{P/C}_{ij} = \mp \frac{\bra{q_iq_j}\hat{H}_2\ket{q_jq_i} - \bra{q_iq_j}\hat{H}_2\ket{q_iq_j}\left| \bra{q_i}\ket{q_j} \right|^2}{1 \mp \left|\bra{q_i}\ket{q_j} \right|^2},
        \label{eq:correlation_potential}
    \end{equation}
    which can be shown to scale as $\left|\bra{q_i}\ket{q_j} \right|^2$ for large particle separations. Due to gaussians being localised states, this gives a short-range interaction between particles $i$ and $j$. The background term $V_{\text{bg}}$ introduces a long-range dependence in terms of the third particle and is accounted for by a type of Ewald summation, see appendix \ref{app:Long-Pauli}. The state averages of the interaction terms in equation \eqref{eq:correlation_potential} are evaluated based on the gaussian mode decomposition. 
    
    The correlation potential, the spin interaction between opposite spin electrons, is constructed based on the same premise as the Pauli potential in a pair-wise approximation, however, with an additional parameter $\rho$ which needs to be chosen a-priori. In the case of a ground state helium or molecular hydrogen, the electronic structure is well described by a single state and $\rho = 1$ is an appropriate choice. For a free electron gas, this would overestimate the correlation effects as the appropriate two-particle state is not simply the singlet state, and therefore $\rho < 1$ is more suitable.
    
    \changed{Lastly, the above-presented schemes -- although widely used -- are in general limited to the weakly and moderately degenerate systems, cause considering the example of Pauli blocking. The Pauli potential in equation \eqref{eq:correlation_potential} appears to be divergent as the orbital overlap tends to unity, however, the numerator vanishes as well when $\ket{q_j} \rightarrow \ket{q_i}$ resulting in only a finite energy barrier. The remaining part of the Pauli exclusion should be accounted for by the left-hand side of equation \eqref{eq:full_wave_packet_dynamics} by a complete anti-symmetrisation scheme.}
        
\subsection{Confining potentials} 
    It has been well documented that at significantly high temperatures wave packets tend to expand indefinitely\,\cite{knaup1999wave,knaup2001wave,knaup2003wave,ebeling2006method,morozov2009localization,davis2020ion} and the wave packet may extend over multiple ions without the ability to localise on multiple sites\,\cite{grabowski2013wave}. If a wave packet is spread too large, it effectively ceases to interact with other particles as the charge density effectively becomes flat. Multiple approaches to counter this expansion has been proposed, see Refs.\,\cite{ebeling2006method,lavrinenko2016reflecting,ma2019extremely}. Currently, we employ an additional potential energy term on the form 
    \begin{equation}
        \mathcal{V}_{\Sigma} = \frac{1}{2}A_{w} \sum_{i = 1}^{N} \sum_{\alpha = 1}^{3} \left( \sigma_{i,\alpha} - l_w \right)^2 \heaviside{ \sigma_{i,\alpha}^2 - l_w^2}
        \label{eq:width_confinment}
    \end{equation}
    where $\sigma_{i,\alpha}^2$ is the $\alpha$th eigenvalue of $\mat{\Sigma}_i$ and $\heaviside{x}$ is the Heaviside step function. The parameters $l_w$ and $A_w$ set the width of a free particle $\sigma_{\text{free}}(l_w, A_w) \gtrsim l_w$ by balancing the shape-kinetic energy. This potential is rotationally invariant and acts only on wave packets with a width larger than $l_w$. Furthermore, the confinement reduces to the commonly used potential based on a harmonic potential centred at the particle position in the limit of $l_w \rightarrow 0$. In this specific limit, the potential has also been used to address the heat capacity in the classical limit\,\cite{ebeling1997quantum}.
%
%\input{text/3_Numerics}
% -------------------- Numerics ----------------------
\section{Numerical Realisation}\label{sec:numerics}
The standard velocity-Verlet integrator almost exclusively used for MD simulations is not appropriate for our model because of the momentum-dependent Pauli potentials\,\cite{feldmeier2000molecular} resulting in a non-separable Hamiltonian. This prevents a straightforward generalisation of the velocity-Verlet algorithm which is based on the ability to separate the Hamiltonian into terms where the dynamics following from each term in isolation can be solved exactly\,\cite{leimkuhler2004simulating}. Explicit Runge-Kutta methods of orders $2$ and $4$ are employed instead. Furthermore, this momentum dependence of the potential affects the interpretation of temperature in the system, further described in appendix~\ref{app:temperature}. 

\begin{figure}
    \centering
    \includegraphics[width=13cm]{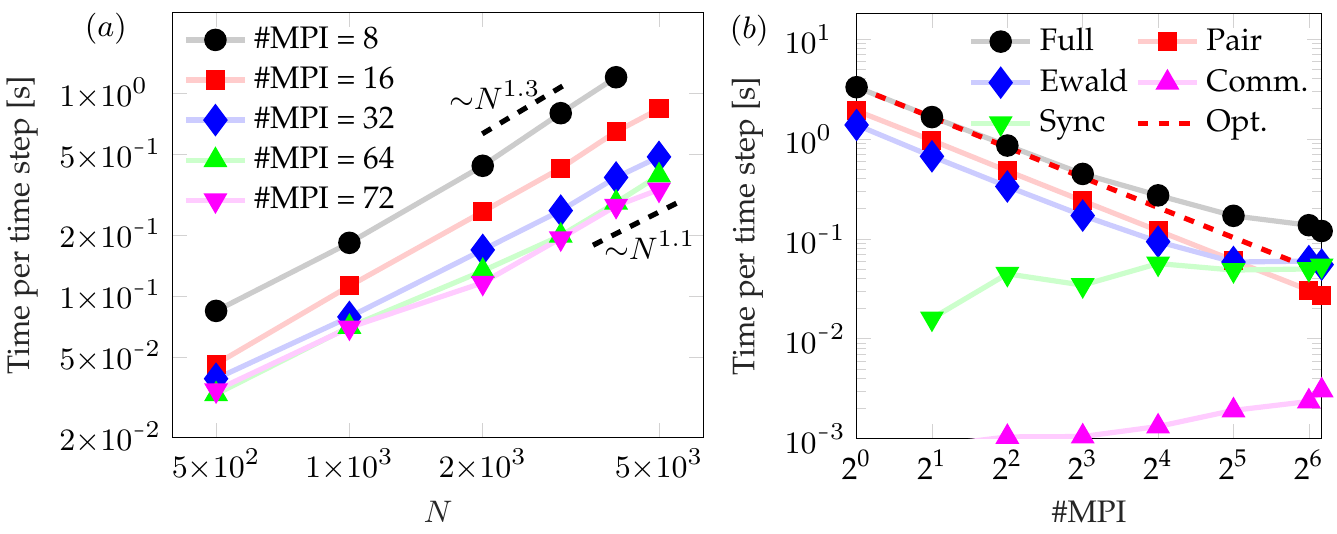}
    \caption{Scaling of computational cost for a quasi-neutral system with $N$ protons, density corresponding to $r_s = 2$ and temperature equal to the Fermi temperature. $(a)$ Scaling with particle number for different levels of parallelisation, in the range between $N^{1.1}$ and $N^{1.3}$. The Ewald parameter $g$ was optimised to within $\pm 0.5a_{B}^{-1}$ for each case. $(b)$ MPI parallelisation of the computation for a system of $N = 2000$ ions with a fixed Ewald parameter. Showing the full computation (circles), pair interaction (squares), Ewald summation (diamonds), communication between MPI processes (upwards triangles), and the synchronisation time different processors need to wait for each other due to an unbalanced load (downwards triangles), which is compared to the optimal scaling based on the single thread performance (dashed).}
    \label{fig:computation}
\end{figure}

The time integrator, the generalised Ewald summation, and the Pauli interaction, are all natively implemented in \texttt{LAMMPS}\,\cite{plimpton1995fast}, a MD framework written in \texttt{C++} which utilises \texttt{MPI} to distribute the computation~\cite{gabriel2004open_mpi}. Figure \ref{fig:computation}$(b)$ shows the computational time for a varying degree of parallelisation for a fixed system of 2000 protons and an equal number of electrons. In particular, a good scaling of the pair interaction and the Ewald summation is established as the computation is distributed. The synchronisation time, the time different processes need to wait for each other due to an unbalanced load caused by statistical fluctuations in the number of particles in the region assigned to each processor, limits the efficiency of the parallelisation when only a few particles are assigned to each process. In the future, dynamic load balancing could potentially address this issue\changed{, however, it should be noted that the point of the plateau moves further out as the size of the system is increased}. 

Furthermore, figure \ref{fig:computation}$(a)$ also demonstrates the scaling of computational cost with particle number for a test system where $r_s = (4\pi/3 \; na_B^3)^{-1/3} = 2$ where $n$ is the proton number density and $a_B$ is the Bohr radius. Close to linear scaling in particle number is demonstrated, showing the feasibility of employing this modeling technique for large systems of particles. The exact exponent varies in the range $1.1 - 1.3$, depending on the degree of parallelisation, caused by different limiting factors in the computation. The synchronisation time is most likely one of these factors for the high parallelisation case. 

%
%\input{text/4_Test}
% ---------------------- Test -----------------------
\section{Test Systems}\label{sec:validation}
    \subsection{Ground state properties}
        Ground state properties of the wave packet models can be obtained by the introduction of a generalised friction term into the equations of motion \eqref{eq:EoM}. Some care is needed to guarantee continuous energy loss due to the momentum dependence of the Pauli potential. This is further described in appendix \ref{app:friction}.  
        
\begin{figure}
            \centering
            \includegraphics[width=8.8cm]{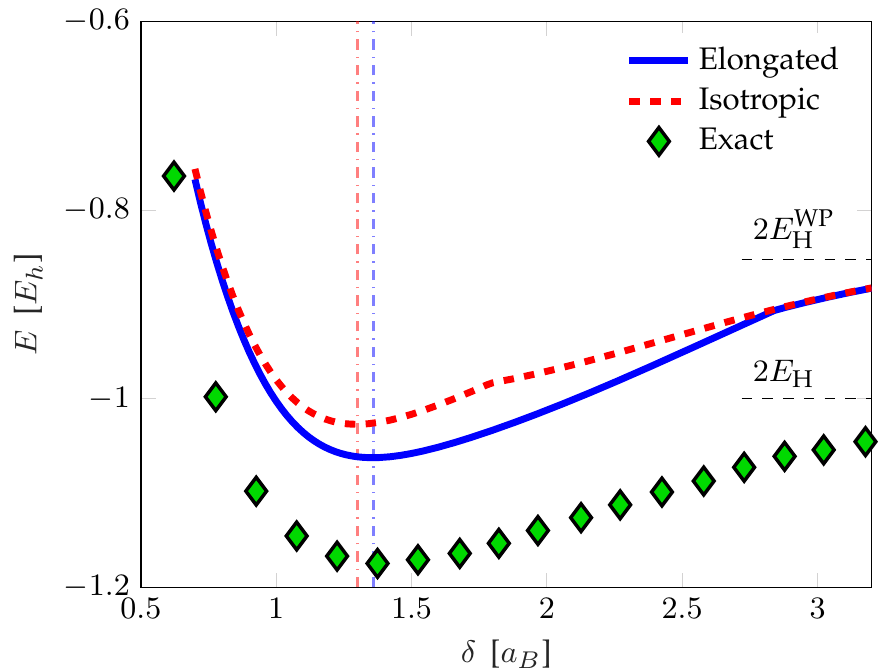}
            \caption{Ground state energy of $\text{H}_2$-molecule, $E$, as a function of nuclear separation $\delta$, computed for the full wave packet model (Elongated), the isotropic one (Isotropic) and the virtually exact result from Ref.\,\cite{angermeier2021investigation} (Exact). The ground state is modeled as a triplet state, with $\rho = 1$ in the correlation potential \eqref{eq:correlation_potential}. The equilibrium displacement is marked for the two wave-packet models (vertical lines) and the energy of two isolated hydrogen atoms for the wave packets as well as the exact result are shown with horizontal markers.}
            \label{fig:H2}
        \end{figure}

        The ground state of isolated atoms is spherically symmetric and does not utilise the additional degrees of freedom of the wave packets. One of the simplest physical systems which naturally breaks this symmetry are diatomic molecules and in particular diatomic hydrogen ($\text{H}_2$), where the ability for the wave packet to stretch is believed to be crucial for molecular binding\,\cite{grabowski2014review}. The ground state energy of $\text{H}_2$ within the wave packet model for both elongated and isotropic gaussians is shown in figure \ref{fig:H2} for a varying nuclear separation $\delta$. The elongated wave-packets demonstrate an improvement over the isotropic ones when $\delta < 2.8a_B$, above which the electron density is localised on each ion and close to spherically symmetric. Furthermore, the isotropic model transitions to an electron density localised on each nucleus at a significantly shorter nuclear separation, $\delta \approx 1.8a_B$, compared to our model at $\delta \approx 2.8a_B$. 

        The energy difference between our model and the exact result is close to constant for nuclear separation larger than the equilibrium position. In this respect, the presented wave packet model is as good a descriptor of the molecular ground state, $\text{H}_2$, as the atomic one, $2\times\text{H}$, suggesting that for further improvements one needs a more versatile description of the isotropic atomistic limit first. The present model has the ability to treat each direction separately and what is limiting the agreement is the restriction on functional form.
        
    \subsection{Binary collisions} % SOFT
    
\begin{figure}
            \centering
            \includegraphics[width=13.5cm]{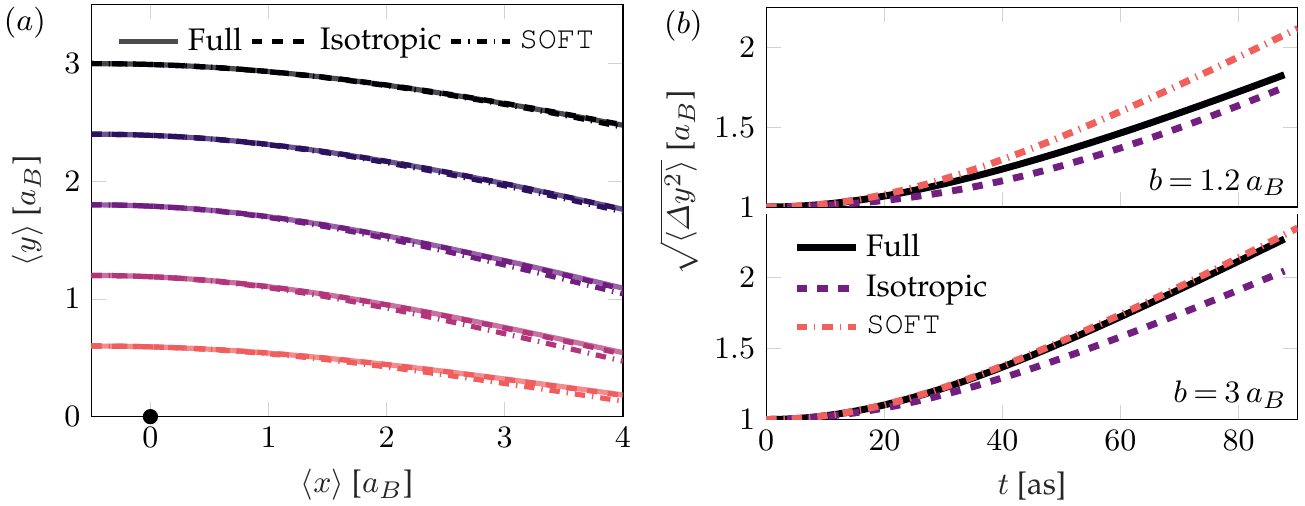}
            \caption{Electron scattering of a fixed proton with different impact parameters $b$ computed using elongated wave packets, isotropic wave packets and \texttt{SOFT}. The trajectory of electron mass-centre $(a)$ and the width of the electron wave function along $y$-direction, $\sqrt{\langle y^2 \rangle - \langle y \rangle^2}$, for two different impact parameters $(b)$. The wave functions were initialised as isotropic gaussians with a width $1\,a_B$ and velocity such that the total classical energy of each trajectory were $\frac{3}{2}k_BT$ and \mbox{$k_BT = 10\,\text{eV}$}.}
            \label{fig:SOFT}
        \end{figure}

\begin{figure}
            \centering
            \includegraphics[width=13.5cm]{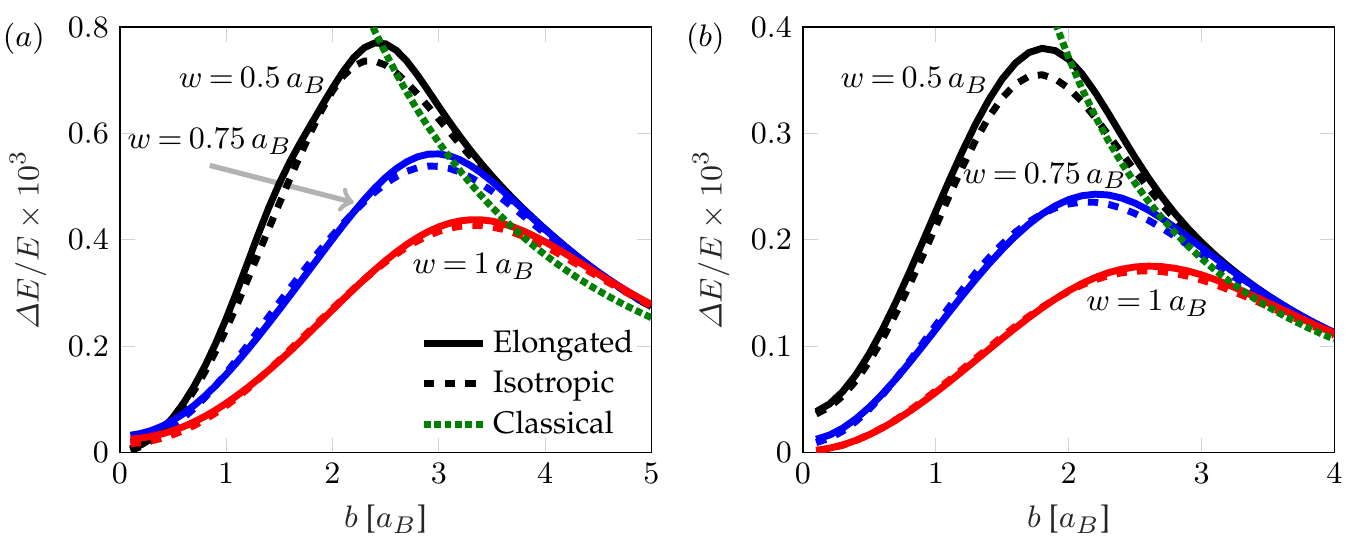}
            \caption{Transfer of (classical) kinetic energy $\Delta E / E$ from the electron to the ion in a single electron-ion scattering event as a function of impact parameter $b$. Initial velocity set as in figure \ref{fig:SOFT} with a temperature $(a)$ $k_{B}T = 5\,\text{eV}$ and $(b)$ $k_{B}T = 10\,\text{eV}$. The energy transferred differs between the isotropic and elongated models, especially for intermediate impact parameters and smaller wave packets. The result is shown for different free particle widths $w$ set by the confining potential in equation \eqref{eq:width_confinment} by varying $l_w$ with $A_w = 3\,E_h/a_B^2$.}
            \label{fig:energy_transfer}
        \end{figure}

        The dynamical properties of the wave packet model have been tested for electron-ion scattering against a full numerical solution of the Schrödinger equation realised by the \texttt{SOFT} code\,\cite{feit1982solution,markmann2012kepler,greene2017tensor}. The resulting trajectory data for impact parameters in the range of $0.6 - 3.0\,a_B$ is shown in figure~\ref{fig:SOFT}, alongside the time evolution of the extent of the wave function, which is compared with the result from both isotropic and elongated wave packets. The centre of mass trajectories agree well between all three different sets of simulations over long-length scales. In the full numerical solution, the electron density can split and partially bind to the ion core, a qualitative feature the wave packets cannot reproduce due to their limited functional form. However, this is of minimal importance for the trajectories discussed here. At these energies, only a minor fraction of the electron wave packet gets bound so that the centre of mass agrees well with the mode position in the full numerical simulation. The two wave packet models differ in the internal degrees of freedom, where the isotropic wave packet does not have the flexibility of the complete model. The isotropic wave packet model cannot reproduce the internal dynamics of the \texttt{SOFT} computation as well as the elongated wave packet model for the larger set of impact parameters. In the case of smaller impact parameters, the full numerical solution has wave packets with non-gaussian structure during the close approach between the electron and the ion, impacting the subsequent evolution of the wave function. 
        One of the more important characteristics for determining the dynamic properties of the system is the energy transfer in particle collisions, which is shown for the electron-proton scattering in figure \ref{fig:energy_transfer}. The energy transfer is calculated by treating the proton dynamically in contrast to the case shown in figure \ref{fig:SOFT}. Differences between the wave packet models can in particular be seen for intermediate impact parameters, between the classical behaviour for large impact parameters and the symmetrical configuration when $b = 0$. Furthermore, the difference is the most pronounced for smaller wave packets where the asymmetry on the scale of the wave packet is more pronounced. The result suggests there might be an appreciable difference in the dynamical properties of the two models.

        \subsection{Transport properties}\label{sec:transport} % More specific

        The previously shown demonstrations of the model were limited to the dynamics of a few particles, however, one of the strengths of the wave packet models is in the treatment of large collections of particles. We demonstrate this by studying a hydrogen plasma with degeneracy parameter $\theta = k_B T / E_{\text{F}} = 1$ and a density $r_s = 2$ under WDM conditions. The system is modeled by a thousand protons and an equal number of electron wave packets describing a spin un-polarised electron fluid (with an equal number of spin-up and spin-down electrons). For this initial test we set $\rho = 1$ in equation \eqref{eq:correlation_potential} and the width of the wave packets are regularised by the width confinement where $A_w = 3\,E_{h}/a_B^2$ and $l_w = 1\,a_{B}$. The initial random configurations of particles were allowed to thermalise under the influence of periodic velocity re-scaling (see appendix \ref{app:temperature}) for $75\,\text{fs}$, after which data was collected for $225\,\text{fs}$, a procedure repeated five times for each wave packet model. 

        \begin{figure}
            \centering
            \includegraphics[width=13cm]{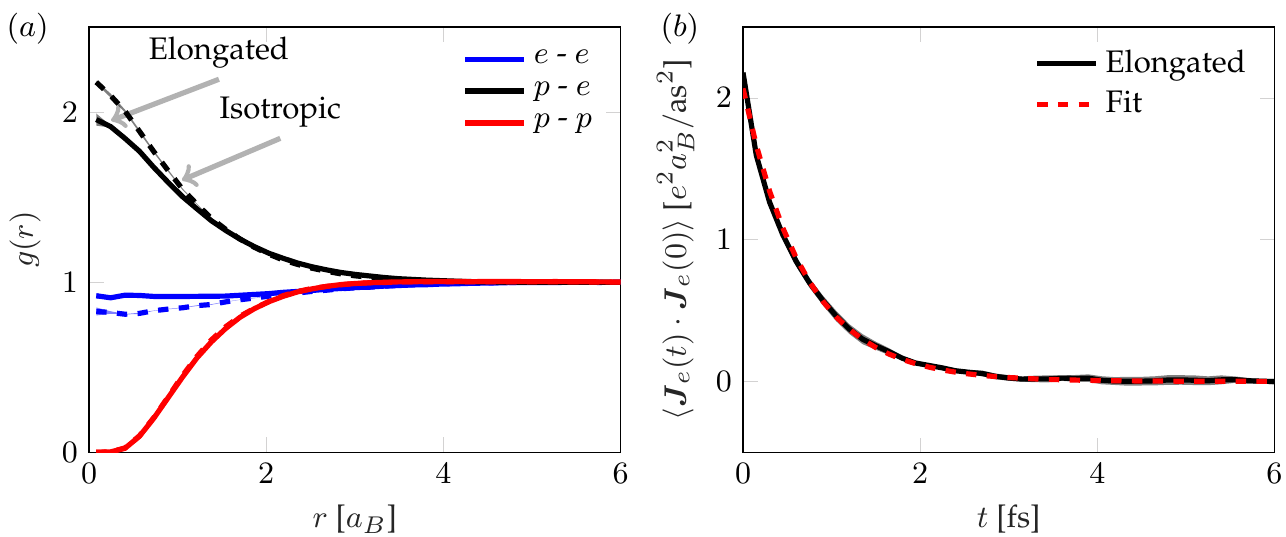}
            \caption{$(a)$ Radial distribution function for proton-proton ($p$-$p$), electron-electron ($e$-$e$) and proton-electron ($p$-$e$) in both wave-packet models described, elongated (solid) and isotropic (dashed). Qualitatively similar structures is seen between the two models, where for isotropic wave packets the electrons has a stronger interaction between themselves as well as the ions. The typical errors -- estimated from the distinct runs -- are shown in the shaded area only noticeable at small $r$. $(b)$ Electron current auto correlation function for the wave packet model (solid), described well by an exponential decay shown by the corresponding fit (dashed). The system considered in both $(a)$ and $(b)$ is the one described in section \ref{sec:validation}\ref{sec:transport}.}
            \label{fig:rdf}
        \end{figure}

        Figure \ref{fig:rdf}$(a)$ shows the static pair-correlation functions~\cite{ichimaru1982strongly} for the two wave packet models under consideration, \changed{computed classically without accounting for anti-symmetrisation of the electrons, and such even equal spin electrons have a contribution at no separation where otherwise the anti-symmetrisation completely set this limit. The static structure} primarily differ in electronic structure. The isotropic wave packets are seen to interact more strongly both with ions and between themselves. A part of this stronger interaction is during a collision between a wave packet and a proton. The isotropic wave function expands slower, as the appropriate expansion is averaged over all directions, and the smaller wave packets have a more localised charge and a stronger interaction.

        One of the properties of interest for WDM systems is the electrical conductivity $\sigma$. The conductivity relates to the microscopic dynamics in an atomistic simulation via\,\cite{vashishta2008interaction}, 
        \begin{equation}
            \sigma(\omega) = \frac{1}{3 k_B\!T\, V} \int_0^{\infty} \tave{\vec{J}(t) \cdot \vec{J}(0)} e^{i\omega t}\,dt,
        \end{equation}
        where $\tave{\cdot}$ is a thermal average and $\vec{J}$ is the total current. The current is dominated by the electron contribution, $\vec{J}_e$, 
        \begin{equation}
            \vec{J} \approx \vec{J}_e = -\frac{e}{m}\sum_{i = 1}^{N_e} \langle \hat{\vec{p}}_i \rangle,
        \end{equation}
        where we sum over all electrons. The current-current correlation function has roughly an exponentially decaying behaviour and therefore in the vein of Ref.\,\cite{mithen2012molecular} an exponential form is fitted to reduce the influence of noise. Both the numerical data and the fit is shown in figure \ref{fig:rdf}$(b)$ with a high level of agreement. An exponential current-current correlation function results in a Drude-like conductivity, 
        \begin{equation}
            \sigma(\omega) = \frac{\sigma_{0}}{1 + \left( \omega/\lambda\right)^2} + i\,\frac{\sigma_{0}\omega / \lambda}{1 + \left( \omega/\lambda\right)^2},
        \end{equation}
        where $\sigma_0$ and $\lambda$ are related to the amplitude and time constant of the decaying correlation function. Physically, $\sigma_0$ is the DC conductivity while $\lambda$ corresponds to the inverse of the mean free scattering time. The wave packet models differ in their prediction of both constants. For elongated wave-packets, $\sigma_0^{\text{elo}} = (34\,300 \pm 1\,900)~\text{\textOmega}^{-1} \text{cm}^{-1}$ and $\lambda^{\text{elo}} = (1.48 \pm 0.06)~\text{fs}^{-1}$, while for the isotropic model $\sigma_0^{\text{iso}} = (29\,800 \pm 1\,200)~\text{\textOmega}^{-1}\text{cm}^{-1}$ and $\lambda^{\text{iso}} = (1.73 \pm 0.06)~\text{fs}^{-1}$. This corresponds roughly to a $15\%$ increase in DC conductivity and a $15\%$ decrease in the scattering frequency as we extend the wave packet formulation. For a Drude-like conductivity, $\sigma_0$ and $\lambda$ are strongly correlated, which can be understood by the constraints set by the fluctuation-dissipation theorem~\cite{kubo1966fluctuation}. Within this wave packet formulation the electron-electron scattering is explicitly included in the dynamic formulation, not the case in the commonly used \mbox{Kubo-Greenwod} formulation of conductivity for DFT-MD, preventing the DFT based technique to achieve the appropriate high temperature limit~\cite{french2022electronic}.
%
%\input{text/5_Result}
%\input{text/6_Conclusion}
% ------------------- Conclusion ------------------
\section{Conclusion}
Non-equilibrium modelling of quantum many-body systems is a formidable task that further suffers from the large proton-to-electron mass ratio, resulting in vastly different time scales for the evolution of electron and ion dynamics. Any dynamical treatment must therefore resolve the electron motion, while extending over comparatively long time scales to investigate ion dynamics. Wave packet models address this with an ansatz for the electronic wave function allowing the time evolution of a large number of particles to be performed over long time scales, while retaining quantum mechanical properties dynamically within the model. 

We have extended the functional form of the wave packets used for modelling WDM by allowing the wave packet to be elongated with arbitrary rotation. This in turn allows for a dynamic response to gradients across the wave packet that should better represent quantum dynamics. As a consequence of the non-isotropic states, explicit evaluation of the interaction Hamiltonian has not been possible, and a generalised Ewald summation has been used to appropriately evaluate both the short- and long-range effect of the Coulomb interaction. Furthermore, a decomposition of the short-range interaction kernel into gaussian modes has been constructed and used for an explicit evaluation scheme.

Crucially, warm dense matter systems are partially degenerate and the exchange interaction contributes significantly to their evolution. A scalable approximation for this in terms of Pauli potentials has been derived as they fundamentally depend on the wave packets used. The interactions are implemented along with the complete dynamical description in \texttt{LAMMPS}, with good parallel support and close to a linear scaling with particle number. 

The elongated wave packet model is seen to improve the description of ground state hydrogen molecules compared to isotropic wave packets where spherical symmetry is naturally broken. The added degrees of freedom also further improve the dynamical description when compared to full quantum mechanical treatments, which we demonstrate for electron-proton scattering. Furthermore, in such collisions different energy transfers are observed between the models, illustrating an important variation in the predicted behaviour. Finally, the model is used to investigate a partially degenerate hydrogen plasma. Differences between wave packet models are seen both in the electronic structure and in the dynamics. A property of fundamental interest for this type of system is the electrical conductivity, which we extracted from our wave packet models. The electrical conductivity is dominated by electron motion and the DC conductivity was seen to increase by approximately $15\%$ when extending the functional form of the wave packet.\vskip6pt

\enlargethispage{20pt}

%\ethics{Insert ethics text here.}
%\dataccess{Insert data access text here.}
%\aucontribute{Insert author contribute text here.}
%\competing{Insert competing text here.}

\noindent\textbf{Founding:} This work was financially supported by AWE UK via Oxford Centre for High Energy Density Science (OxCHEDS) scholarship jointly founded by Oxford Physics Endowment for Graduates (OxPEG). \changed{ZM gratefully acknowledges funding by the Center for Advanced Systems Understanding (CASUS) which is financed by Germany’s Federal Ministry of Education and Research (BMBF) and by the Saxon state government out of the State budget approved by the Saxon State Parliament. V.S.B.~acknowledges support from National Science Foundation Grant CHE-1900160 for the calculations performed by N.L}

\noindent\textbf{Acknowledgement:}The authors would like to thank S.\ Azadi and T.\ Gawne for valuable discussions.

%\disclaimer{Insert disclaimer text here.}
\appendix
%\input{text/A_Decomposition}
% -------------------- Decomposition ------------------
\section{Gaussian decomposition}\label{app:decomposition}
The decomposition of the Coulomb interaction kernels described in section \ref{sec:theory}\ref{sec:ewald} is solely introduced to efficiently evaluate state averages and only needed to be computed once. The state averages of the interaction kernel $V$ can be reduced to a volume integration weighted by some gaussian profile. Therefore, the error in gaussian approximation was quantified by the volume average of the square errors
\begin{equation}
    \mathcal{L} = \int_{0}^{r_{\text{max}}}\!\!dr\, r^2 \Big|V(r) - \sum_p c_p e^{-\alpha_p r^2}\Big|^2
\end{equation}
which is finite without a small $r$ cutoff. In the case of the short-range contribution, \mbox{$V(r) = \erfc\left(gr\right)/r$}, the integrals are convergent and $r_{\text{max}} = \infty$ was chosen. On the other hand for the long-range correction, $V(r) = \erf\left(gr\right)/r$ a finite $r_{\text{max}}$ is needed and has been chosen to $r_{\text{max}} = 15a_B$. \changed{The latter decomposition is only applied to terms in the Pauli interaction, see section \ref{sec:theory}\ref{sec:pauli}, which are exponentially suppressed by the overlap $\left|\bra{q_i}\ket{q_j} \right|^2$ for large distances and the self-energy for the Ewald summation which is also exponentially suppressed for large distances, see appendix \ref{app:ewald}. Therefore a finite $r_{\text{max}}$ does not constitute a significant limitation.} 

\begin{figure}
    \centering
    \includegraphics[width=13cm]{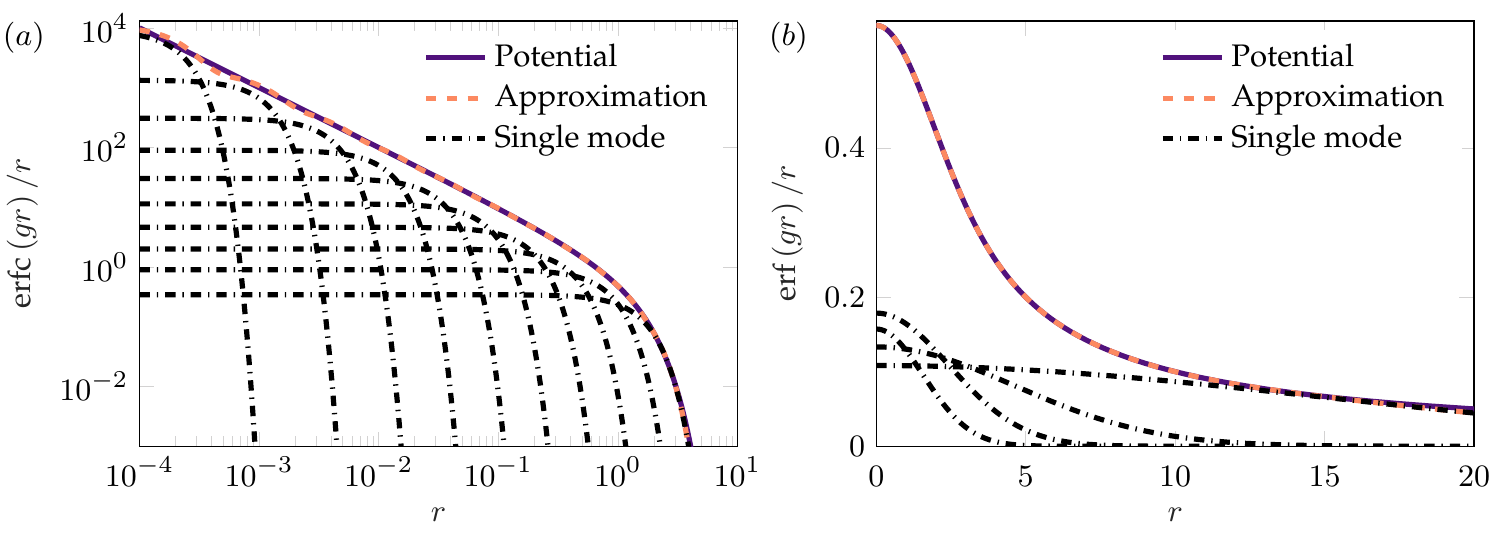}
    \caption{The gaussian mode decomposition of the short-range interaction $(a)$ and the long range-interaction $(b)$ with a Ewald parameter $g = 0.5$.}
    \label{fig:decomposition}
\end{figure}

The decomposition in terms of the coefficients $c_p$ and $\alpha_p$ is based on the minimisation of $\mathcal{L}$. For a fixed set of $\alpha_p$, the minimisation is solved by inversion of the linear system of equations ${\partial \mathcal{L} / \partial c_p = 0}$. However, instead of using a particular set of $\alpha_p$\,\cite{shigeta2013quantal} a general minimisation algorithm was used to find the optimal $\alpha_p$'s, preconditioned with the optimal $c_p = c_p(\left\{\alpha_p\right\})$. For the short-range interaction the search was limited to ${\alpha_p > 0}$ where in the long-range case ${r_{\text{max}}^{-2}/2 < \alpha_p < g^2}$ was needed for a stable decomposition. Figure \ref{fig:decomposition} demonstrates the final decomposition resulting from this procedure described above, typical for the ones used in the main text. The singular nature of the short-range interaction is approximated by a series of modes with successively larger amplitude and shorter range. 

%\input{text/B_Ewald}
% ---------------------- Ewald ----------------------
\section{Ewald summation}\label{app:ewald}

The main contribution $\mathcal{E}_k$ to the Ewald summation \eqref{eq:ewald_summation} is the interaction of the charge distribution in the simulation box $\rho_{\text{uc}}$ and the total surrounding charge distribution $\rho_{\text{tot}}$,
\begin{equation}
    \mathcal{E}_k = \int d^3\vec{x}_1 d^3\vec{x}_2 \frac{\erf\left( g |\vec{x}_1 - \vec{x}_2| \right)}{2|\vec{x}_1 - \vec{x}_2|}\rho_{\text{uc}}\left( \vec{x}_1 \right) \rho_{\text{tot}}\left( \vec{x}_2 \right), 
\end{equation}
seen by exchanging the particle summation and state average in equation \eqref{eq:ewald_summation}. The charge distributions are based on the functional form of the wave packet and the image charges, namely
\begin{equation}
    \rho_{\text{uc}}\left(\vec{x}\right) = \sum_{i} e Z_i |\bra{\vec{x}}\ket{q_i}|^2 \qquad \text{and} \qquad \rho_{\text{tot}}\left( \vec{x} \right) = \sum_{\vec{n} \in \mathbb{Z}^3} \rho_{uc}\left(\vec{x} - L\vec{n}\right),
\end{equation}
the Fourier transform of which has been evaluated in equation \eqref{eq:FT_density}. With the use of Plancherel’s theorem and the Fourier convolution theorem, one can arrive at the expression \eqref{eq:long_main} which converges rapidly in $K$-space due to the exponential dependence on $\vec{k}$. The $\vec{k} = 0$ term vanish due to charge neutrality $\Tilde{\rho}_{\text{uc}}(0) = 0$. 

In any given computation only a finite number of $\vec{k}$-vector may be used and the associated error incurred by truncating equation \eqref{eq:long_main} was described for the classical point particle case by Ref.\,\cite{kolafa1992cutoff}. It can be shown that the resulting force from the Ewald summation for distributed particles are suppressed by factors $\exp\left(- 0.5\vec{k}^{\intercal}\mat{\Sigma}_i\vec{k}\right) \times \exp\left(- 0.5\vec{k}^{\intercal}\mat{\Sigma}_j\vec{k}\right)$ for each $\vec{k}$-vector and the force converge more rapidly -- as the charge distribution now is more smooth -- such that the classical estimate still can be used as an upper bound. The calculations in the manuscript were performed such that this error estimate was one thousand of the force two ions experience an average distance apart.     
    
The self-energy term can be formulated similarly:
\begin{equation}
    \mathcal{E}_s = -\sum_{i}\left(eZ_i\right)^2 \!\int \! d^3\!\vec{x}_1 |\bra{\vec{x}_1}\ket{q_i}|^2 \int d^3\!\vec{x}_2\, \frac{\erf\left( g |\vec{x}_1 - \vec{x}_2| \right)}{2|\vec{x}_1 - \vec{x}_2|} |\bra{\vec{x}_2}\ket{q_i}|^2,
\end{equation}
where the integration could be performed if the kernel were gaussian. The integrand is exponentially small if $\vec{x}_1 \not\approx \vec{x}_2$ and a gaussian decomposition of the long-range interaction kernel is utilised, using $\erf\left( gr \right) / r = \sum_{p} \Tilde{c}_p \exp\left(-\Tilde{\alpha}_p r^2 \right)$. The self-energy contribution is 
\begin{equation}
    \mathcal{E}_s = - \frac{1}{2}\sum_{i} \left(eZ_i\right)^2 \sum_{p} \frac{\Tilde{c}_p}{\sqrt{\det\left( \mat{I} + 4\Tilde{\alpha}_p\mat{\Sigma}_i \right)}}, 
\end{equation}
yielding the correct classical limit, $\mat{\Sigma}_i \rightarrow 0$, as $\sum_p \Tilde{c}_p = 2g/\sqrt{\pi}$, if the decomposition retain the correct value at $r = 0$. 
%
%\input{text/C_Ewald_Pauli}
% ----------------------- Ewald_Pauli -----------------
\section{Log-range Pauli interaction}\label{app:Long-Pauli}
The background terms, $V_{\text{bg}}$, in the Pauli potential \eqref{eq:pauli_bg} have a long-range behaviour and by splitting the interaction in accordance with the Ewald summation discussed in section \ref{sec:theory}\ref{sec:ewald} it may be truncated and any residual long-range term is then accounted. Once again the long-range interaction can be formulated as the interaction between two distributions; one is the particle overlap and other is the charge density from the background particles: 
\begin{equation}
    \mathcal{V}^{P/C}_{\text{long}} = \int d^3\vec{x}_1 d^3\vec{x}_2 \frac{\erf\left( g |\vec{x}_1 - \vec{x}_2| \right)}{2|\vec{x}_1 - \vec{x}_2|} \mu^{P/C}(\vec{x}_1) \rho_{\text{tot}}(\vec{x}_2) + \mathcal{V}_s^{P/C}
\end{equation}
where 
\begin{equation}
\begin{aligned}
    \mu^{P/C}(\vec{x}) = \pm\sum_{i < j} \frac{e}{1 \mp |\bra{q_i}\ket{q_j}|^2} \Big[ 2 &\Re\left\{ \bra{q_j}\ket{\vec{x}}\bra{\vec{x}}\ket{q_j}\bra{q_j}\ket{q_j} \right\}\\
    &- \left( |\bra{\vec{x}}\ket{q_i}|^2 + |\bra{\vec{x}}\ket{q_j}|^2 \right)|\bra{q_i}\ket{q_j}|^2  \Big]
\end{aligned}
\end{equation}
and the sum is restricted to electrons with relevant spins.  A self-energy term $\mathcal{V}_s^{P/C}$ is introduced when the background terms do not have terms where $k = i, j$, 
\begin{equation}
\begin{aligned}
    \mathcal{V}_s^{P/C} = \pm\sum_{i < j} \int &d^3\vec{x}_1 d^3\vec{x}_2 \frac{\erf\left( g |\vec{x}_1 - \vec{x}_2| \right)}{2|\vec{x}_1 - \vec{x}_2|} \frac{e}{1 \mp |\bra{q_i}\ket{q_j}|^2} \left( |\bra{\vec{x}}\ket{q_i}|^2 + |\bra{\vec{x}}\ket{q_j}|^2 \right)\\
    &\times \left[ 2 \Re\left\{ \bra{q_j}\ket{\vec{x}}\bra{\vec{x}}\ket{q_j}\bra{q_j}\ket{q_j} \right\} - \left( |\bra{\vec{x}}\ket{q_i}|^2 + |\bra{\vec{x}}\ket{q_j}|^2 \right)|\bra{q_i}\ket{q_j}|^2  \right]
\end{aligned}
\end{equation}
while the above expression has been evaluated in K-space and included in the force computation, it has only a limited impact on the overall dynamics, and it could be omitted if computational speed is needed. 
%
%\input{text/D_Temperature}
% -------------------- Temperature ----------------------
\section{Temperature measurements}\label{app:temperature}
The system temperature, $T$, is defined through the equipartition theorem:
\begin{equation}
    \tave{Q_{\mu}\deri{\mathcal{H}}{Q_{\nu}}} = \delta_{\mu\nu}k_{B}T, 
\end{equation}
where $\delta_{\mu\nu}$ is the Kronecker delta and $\tave{ \cdot }$ represent the thermal average in the microcanonical ensemble commonly referred to as the NVE ensemble, due to a constant particle number, $N$, volume, $V$, and energy $E$. This corresponds to a time average in molecular dynamics in accordance with the ergodic hypothesis\,\cite{rapaport2004art}. We may define a temperature based on the classical degrees of freedom $Q_{\mu} = Q_{\nu} = \vec{p}_i$, 
\begin{equation}
    \tave{\frac{\vec{p}_i^2}{m} + \vec{p}_i^{\intercal} \deri{\mathcal{V}}{\vec{p}_i}} = 3k_{B}T,
    \label{eq:calssical_temp}
\end{equation}
and the internal ones, 
\begin{equation}
    \tave{ \frac{4}{m}\Tr\big\{\mat{\Pi}_i \mat{\Sigma}_i \mat{\Pi}_i\big\} + \Tr\left\{\mat{\Pi}_i \left( \mat{\tau} \odot \deri{\mathcal{V}}{\mat{\Pi}_i} \right)\right\}} = 6 k_{B}T 
    \label{eq:internal_temp}
\end{equation}
where $\mathcal{V}$ is the state average of the interaction terms of the hamiltonian. The above two definitions are combined, averaged over the particle index $i$ and evaluated instantaneously to monitor the temperature evolution of the system. Instantaneously, equations \eqref{eq:calssical_temp} and \eqref{eq:internal_temp} may differ, nonetheless converges to the same value by time averaging. Previously, Ref.\,\cite{ma2019extremely} has incorporated the width kinetic energy in the temperature measurement for isotropic wave packets. When we employ isotropic wave packets we use equation \eqref{eq:internal_temp} accounting for the reduced degrees of freedom.
%\input{text/E_Friction}
% --------------------- Friction ---------------
\section{Generalised friction}\label{app:friction}
To achieve a continuous energy loss from friction, the equations of motion are modified accordingly: 
\begin{equation}
    \frac{d\vec{p}_i}{dt} = - \deri{\mathcal{H}}{\vec{r}_i} + \vec{\Gamma}_i^{p}, \qquad \text{and} \qquad \frac{d\mat{\Pi}_i}{dt} = - \mat{\tau} \odot \deri{\mathcal{H}}{\mat{\Sigma}_i} + \mat{\Gamma}_i^{\Pi}.
\end{equation}
The resulting energy loss 
\begin{equation}
    \frac{d}{dt}\mathcal{H} = - \sum_i \deri{\mathcal{H}}{\vec{p}_i^{\intercal}} \;\vec{\Gamma}_i^{p} + \Tr\left\{\mat{\tau} \odot \deri{\mathcal{H}}{\Pi_i} \mat{\Gamma}_i^{\Pi} \right\}, 
\end{equation}
can be guaranteed to result in a monotonic energy decrease with the choice
\begin{equation}
    \vec{\Gamma}_{i}^{p} = m \gamma_p\; \deri{\mathcal{H}}{\vec{p}_i} \hspace{0.5cm} \mat{\Gamma}_{\Pi} = \frac{m}{8} \gamma_{\Pi} \left( \mat{\Sigma}_i^{-1} \mat{\tau} \odot \deri{\mathcal{H}}{\Pi_i} + \mat{\tau} \odot \deri{\mathcal{H}}{\Pi_i} \mat{\Sigma}_i^{-1}\right) 
\end{equation}
which reduces the classical case in the absence of a Pauli potential and $\mat{\Sigma}_i \mat{\Pi}_i = \mat{\Pi}_i \mat{\Sigma}_i$. 

\printbibliography

\end{document}